# Microwave Amplitude Reflecting Instability of LFP Electrode Ground Field Are Useful for Consciousness State Identification

Chika Koyama*, Taichi Haruna, and Kazuto Yamashita

***Abstract*— *Goal:* We recently developed EEG indices that correlated with anesthesia concentration in dogs by focusing on microwaves of flattish period (named $\tau$). Aims of this study are to discriminate conscious states of mice using our original indices on LFP and to explore the principle of $\tau$. *Methods:* We analyzed 20h LFP with various sampling rates at layer 5 of secondary motor cortex (M2) and primary somatosensory cortex (S1) in 6 mice. $\tau$ is defined as a subthreshold wave and burst wave is defined as an above-threshold wave. For every 60s LFP, the number of $\tau$ ($N\tau$), duration of $\tau$ ($M\tau$), and amplitude of burst (Abst) were computed with various thresholds. *Results:* Changes of new indices against changes of threshold showed the same pattern in all cases, whereas the results of frequency analysis varied. Those indices well quantified the morphological features of waveforms in milliseconds and made it possible to discriminate the specific waveforms of the state of consciousness; awake, shallow sleep, rapid eye movement (REM) sleep, and non-rapid eye movement (NREM) sleep. On the other hand, microwaves often fluctuated for each sample at any rate. $N\tau$ reached the maximum when $M\tau$ was about 2.5 to 3.0 sampling intervals in any state, and that threshold decreased in state order of awake, shallow sleep, REM sleep, and NREM sleep ($p< 10^{-5}$). Threshold differences between NREM and REM sleep states were more pronounce in S1 than M2. *Conclusion:* Microwaves were local fluctuations of electrode, whose amplitude correlated to level of consciousness.***

## I. Introduction

SUPPLY of non-invasive and continuously monitorable objective indicators to discriminate state of consciousness will be a major challenge in a wide range of research and clinical fields. Many studies have shown that neural activity of cortical pyramidal neurons in layer 5 (PnL5) play a core role in the mechanism for generating consciousness [1]-[8]. In particular, the activation of apical dendrites in PnL5 was shown to correlate with the threshold for transition from an insensate to a perceptual state [9], [10]. These studies demonstrated that the generation of perceptions is dependent on apical amplification of PnL5 (AA). They also strongly supported previous reports that adrenergic arousal enhances AA [11]-[13].

Electroencephalogram (EEG) recording does not require a large machine and is now widely used for research and medical fields. However, the current EEG devices have not been accurate enough to prevent intraoperative awareness during anesthesia [14],[15]. Recent researches using multichannel EEG focusing on functional connectivity of brain circuits have also shown that these connectivity patterns are unsuitable for assessing level of consciousness because of their instability [16], [17].

We have recently developed new EEG indices focusing on microwaves, and reported that these linearly correlated with volatile anesthesia concentration in dogs [18]. EEG is generated mainly by postsynaptic potentials of PnL5 and contains flattish periods where there is not much variance in voltage between successive peaks in any state of consciousness. Those sub-threshold waves are defined as "**$\tau$**" periods and the above-threshold waves (waves between two adjacent **$\tau$**s) are defined as "**burst**" periods (ex. Fig.1d-f). In this study, we demonstrate that these new indices from local field potential (LFP) of PnL5 in mice were also used robustly to discriminate between state of consciousness, together with the principle of **$\tau$**. First, we noted that the change with increasing the sampling frequency of LFP highlights differences in waveforms between awake and sleep states (Fig.1). Compared with the 125Hz waveforms, the 1000Hz waveform have new spike-like waves of various voltages in awake state (Fig.1d), whereas the waves become finer in sleep state (Fig.1ef). These are consistent with previous reports that arousal enhances AA [11]-[13]. Next, assuming that AA is always quiescent throughout sleep, we hypothesize the following three based on observations from Fig.1; 1) When the threshold is around 0.1mV or higher, number of **$\tau$** will increase in awake state and will decrease in sleep state as sampling rate of data increases. 2) Conversely, for small threshold such as 0.03mV or less where fragmented microwaves are detected as a **$\tau$**, the number of **$\tau$** in sleep will be greater than that in awake. 3) Mean amplitude of burst will be smaller in sleep than in awake.

This paragraph of the first footnote will contain the date on which you submitted your paper for review.

**Competing Interest:** No external funding and no competing interests declared.

T. H. is with Department of Information and Sciences, Tokyo Woman's Christian University, Suginami, Tokyo, Japan. K. Y. is with Department of Small Animal Clinical Sciences, Rakuno Gakuen University, Ebetsu, Hokkaido, Japan. *C. K. is with Department of Small Animal Clinical Sciences, Rakuno Gakuen University, Ebetsu, Hokkaido, Japan and also with Laboratory for Haptic Perception and Cognitive Physiology, Center for Brain Science, RIKEN, Wako, Saitama, Japan (correspondence e-mail: s21641008@g.rakuno.ac.jp).

In this study, number of **τ** (N**τ**), mean duration of **τ** (M**τ**), and mean amplitude of **burst** (Abst) with various thresholds on 60 s LFP data for 20 h sampled at 125Hz, 250Hz, 500Hz, and 1000Hz at secondary motor cortex (M2) and primary somatosensory cortex (S1) were computed. By demonstrating the robustness in discriminating states of consciousness (awake, shallow sleep, rapid eye movement (REM) sleep, and non-rapid eye movement (NREM) sleep), we propose that, as the principle of **τ**, amplitude of microwaves will indicate fluctuation of the electrode field caused by brain activity. This is the first time to analyze those indices in animals other than dogs, and to explore the morphological changes of microwaves of PnL5 LFP by state of consciousness.

All animal experiments were performed in accordance with institutional guidelines and were approved by the Animal Experiment Committee of the RIKEN Brain Science Institute.

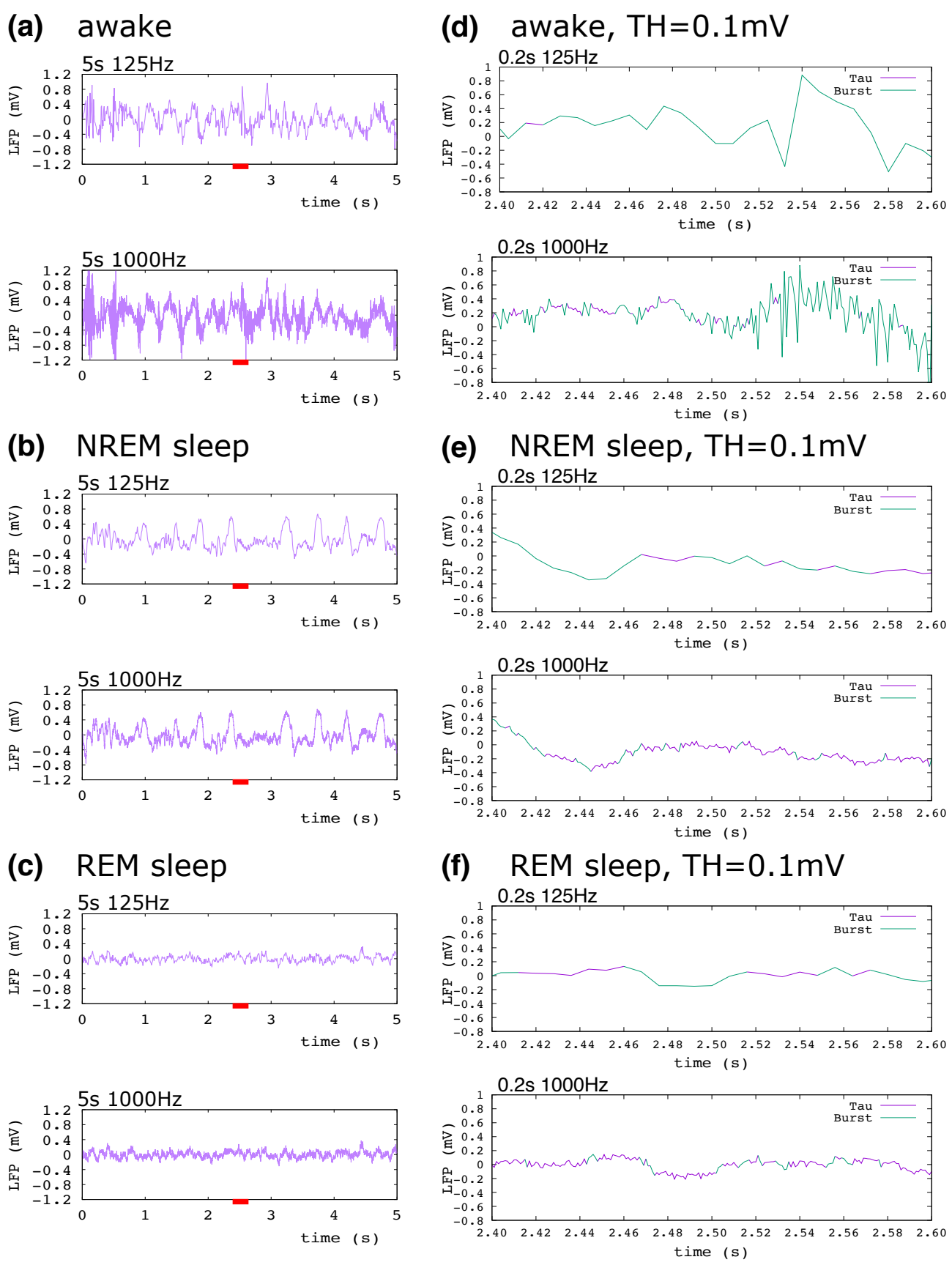

Fig. 1. Typical state-specific waveforms and examples of τ and burst at 0.1 mV threshold (TH), on LFP sampled at 125Hz and 1000Hz at layer 5 of secondary motor cortex. (a-c): Representative 5 s LFP waveforms during awake (a), NREM sleep (b), and REM sleep (c), sampling rate of 125Hz (top) and 1000Hz (bottom). (d-f) Example of **τ** and **burst** in enlarged views between 2.4 s and 2.6 s of the 5 s LFP at 125Hz (top)and 1000Hz (bottom) during awake (d), NREM sleep (e), and REM sleep (f). The **τ** periods below TH of 0.1mV are colored purple and **burst** periods above TH of 0.1 mV are colored green.

## II. Material and Methods

### A. Subjects

Six Thy1-ChR2 (Jackson Laboratory, Bar Harbor. ME, USA) maintained on a C57BL/6J background and CA1 specific Cre mouse line (CaMKIIα-Cre;CW2) mice [19] were used. In all experiments, mice were housed in a 12 h-light:12 h-dark (light on: 8 AM/light off :8 pm) cycle in individually ventilated cages with 1or 2 animals per cage.

### B. Data acquisition

Continuous EMG recordings were performed through a slip ring. Continuous LFPs recordings were performed using 75µm platinum electrode from M2 and S1 in the right hemisphere. For targeting cortical layer 5, M2 and S1 electrodes were inserted 670 µm in depth. Those were recorded 24 hours a day from 0 AM. Electrical signals were filtered at from 0.1 Hz to 5 kHz, amplifier and digitized at 10 kHz. LFP at M2 and EMG were used for computer-based online sleep scoring by existed method [20] [21]. LFP data at M2 and S1 were used to examine **τ** and **burst**.

LFP data measured experiments for which detailed protocols have already been reported [22] were used in this study.

### C. Data analysis

The 1200 LFPs data of 60 s were extracted for analysis from the data of 20 hours, excluding the 2 hours each after the start and before the end of recording. Data determined to be the same state for all 60 s were used for **τ** and **burst** analysis and frequency analysis.

### D. *τ and burst analysis*

Peaks of LFP waves were detected through its first-order derivatives. To test the hypothesis, 13 potential difference thresholds were set at intervals of 0.005 from 0.005 to 0.025 (mV), at intervals of 0.025 from 0.025 to 0.150 (mV), and at intervals of 0.050 from 0.150 to 0.300 (mV). To search for comparable conditions, threshold was set every 0.001 mV from 0.001mV to 0.100mV. A subthreshold period where the voltage difference between adjacent peaks does not exceed the threshold is defined as a **τ**, and while an above-threshold period where it exceeds the threshold is defined as a **burst**. Based on **τ** and **burst** components in 60 s of LFP data, we developed the following 3 LFP parameters as following:

**τ: subthreshold period**

**Nτ = total number of τ**

**Mτ (s) = Σ τ (s) / Nτ**

**burst: above-threshold period**

**Abst (mV) = Σ amp (mV) / Nbst**

Σ **τ** (s) denotes the sum of all **τ** events. Nbst is the total number of **burst** events occurring in 60 s and is equal to N**τ** or N**τ** ± 1. Σ amp (mV) is the sum of the difference between the maximum and minimum voltages in a **burst**, which we call **burst** amplitude. M**τ** (s) is mean **τ** duration. Abst (mV) is mean **burst** amplitude. Data were analysed by program written in C.

### E. Frequency analysis

One-second moving averages were subtracted from the data, which were divided into mutually non-overlapping 2 s periods, and Welch's window function was applied. Subsequently, discrete Fourier transform was performed at 50 ± 1 Hz and 100 ± 1 Hz to remove noise caused by the alternating current power source. The power spectrum density was estimated by fast Fourier transform and spectral edge frequency 95 (SEF95) was derived. SEF95 is defined as the frequency below which 95% of the signal power resides. Data were analysed by program written in C.

### F. *Statistical analysis*

We performed agglomerative hierarchical clustering for Abst and Nτ. First, we computed the Euclidean distances as the dissimilarity values by "dist" function, and then feed these values into "hclust" function with Ward's minimum variance method "ward.D2". Next, we cut the dendrogram into 2 groups (awake and sleep) with "cutree" function. The sleep group in Nτ was further classified into 3 groups (NREM, REM, and shallow sleeps) by "hclust" function with "centroid" or "average" method followed by "cutree" function.

To determine a threshold for maximizing Nτ (THnt), the indices were calculated at every 1 μV up to threshold of 100 μV for LFP data. Mean values of indices in each state were calculated in each mouse (Supplementary Figs. 2-5). The mean values of Nτ for threshold were used to determine THnt. The mean value of index at THnt was used as the representative value at state of consciousness in individual. Because analysis data with 4 different sampling frequencies were generated for M2 and S1 LFPs, 8 representative values were set for each state of the individual per a single index for statistical examination.

Bonferroni corrected Wilcoxon signed rank test was used to investigate whether new indices varies with sampling rate of LFP data. Bonferroni corrected Wilcoxon rank sum test was used to compare SEF95 or new indices among 4 states of consciousness. Wilcoxon signed rank test was used to compare indices among M2 and S1. Holm corrected exact Wilcoxon signed rank test was used to compare new indices at THnt among 4 states of consciousness.

Values are expressed as mean ± standard deviation or median (1st quartile, 3rd quartile). All statistical analysis was performed using R (version 3.6.2). Graphical outputs were generated using gnuplot (version 5.1).

## III. RESULTS

Changes of indices against changes of threshold or sampling frequency showed the same pattern in both M2 and S1 of all mice (Nos.1-6). As those summaries, the mean values of Nτ, Mτ, and Abst for each consciousness state at threshold from 0.001mV to 0.100mV are shown in Supplementary Figs. 2-5.

Consistent with our hypothesis, scatterplots of Nτ and Abst, results of indices for 20 h data for each threshold, formed 2 populations; awake group with large variability and sleep group with small variability (Figs.2-4). In Nos.1 and 2, these 2 populations are almost perfect agreement with the state discrimination by the existing method. On 125-250 Hz LFP, the sleep population of Nτ at threshold of around 0.025mV and 0.100-0.150mV, further formed 2 or 3 subpopulations (Figs.3a, 6ab). First, using Abst on 1000Hz LFP at S1, hierarchical clustering was performed for 2-state (awake and sleep) discrimination, and its accuracy, which depends on the presumption that AA is always quiescent throughout sleep, was confirmed by Nτ. Next, using Nτ on 125Hz LFP at M2, hierarchical clustering was performed in two steps for 4-state (awake and, shallow, NREM, REM sleeps) discrimination. Those results are summarized in Fig.6c. Finally, we examined the principle of τ (Figs.7,8).

### A. *Representative change pattern of Nτ at a threshold of 0.100 mV with changes in sampling rate of LFP*

To confirm the first hypothesis, new indices were calculated using LFP for 4 different frequencies (125Hz-1000Hz). As shown in one example using the results of No.1 (Fig.2, Supplementary table1), the changes in Nτ, Mτ, and Abst with increasing sampling frequency are as follows; In awake state, Nτ increased and Mτ decreased. In sleep state, Nτ increased from 125Hz to 250Hz and decreased from 250Hz to 1000Hz and Mτ increased. Abst remained almost unchanged in awake state and decreased in sleep state. Those changed significantly with changes in sampling frequency ($p<10^{-15}$). There is a transient increase in Nτ of the sleep state from 125Hz to 250Hz, which was due to onset of small waves, not spike. The reason for this is that Mτ increased in all cases, and Nτ decreased more at 500Hz and 1000Hz than at 125Hz. The changes in these indices are consistent with the first hypothesis and following events: with increasing sampling frequency, spikes were expressed during awake, whereas waves became finer during sleep (Fig.1).

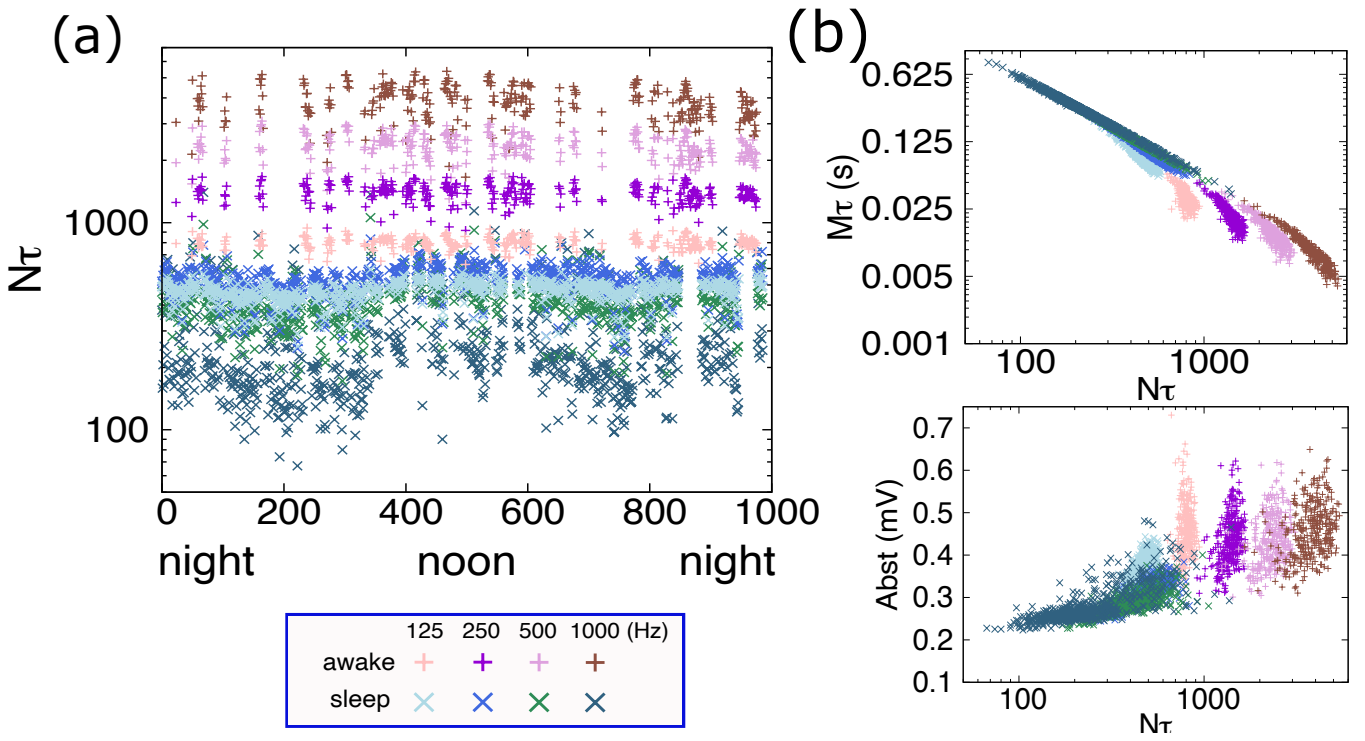

Fig. 2. Example of changes in Nτ for trial number (a), Mτ for Nτ (b), Abst for Nτ (c) at 0.1mV threshold with changes in sampling frequency (No.1). Sampling rate of 125Hz, 250Hz, 500Hz, 1000Hz in LFP are color coded by +, +, +, + in awake state (n = 307), and by ×, ×, ×, × in sleep state (n = 684), respectively.

### B. *Change patterns of Nτ and Abst in awake and sleep states with changes in threshold*

To examine our second and third hypotheses, new indices were examined with various thresholds. The changes in Nτ and Abst with increasing threshold are as follows for awake and sleep states; Nτ increased and then decreased, respectively. Abst decreased and then increased, respectively. In one example using the results of No.1 on 250Hz LFP (Fig.3a), when threshold rises from 0.005mV to 0.025mV, Nτ increases noticeably, while Nτ in sleep state remains larger than Nτ in awake state. At 0.050mV threshold, the difference between Nτ values in both states narrows. When threshold exceeds 0.050mV, Nτ in sleep state decreases noticeably, whereas Nτ in awake state decreases slowly. Here, Nτ in awake state becomes larger than Nτ in sleep state in reverse. Consistent with the second hypotheses, at small thresholds where microwaves are detected separately as τ, Nτ was larger in sleep state than in awake state. At that small threshold with higher frequencies of 500Hz and 1000Hz, the difference in Abst between awake and sleep states was more apparent, as confirmed by visual inspection of the scatterplots (Fig.3bc).

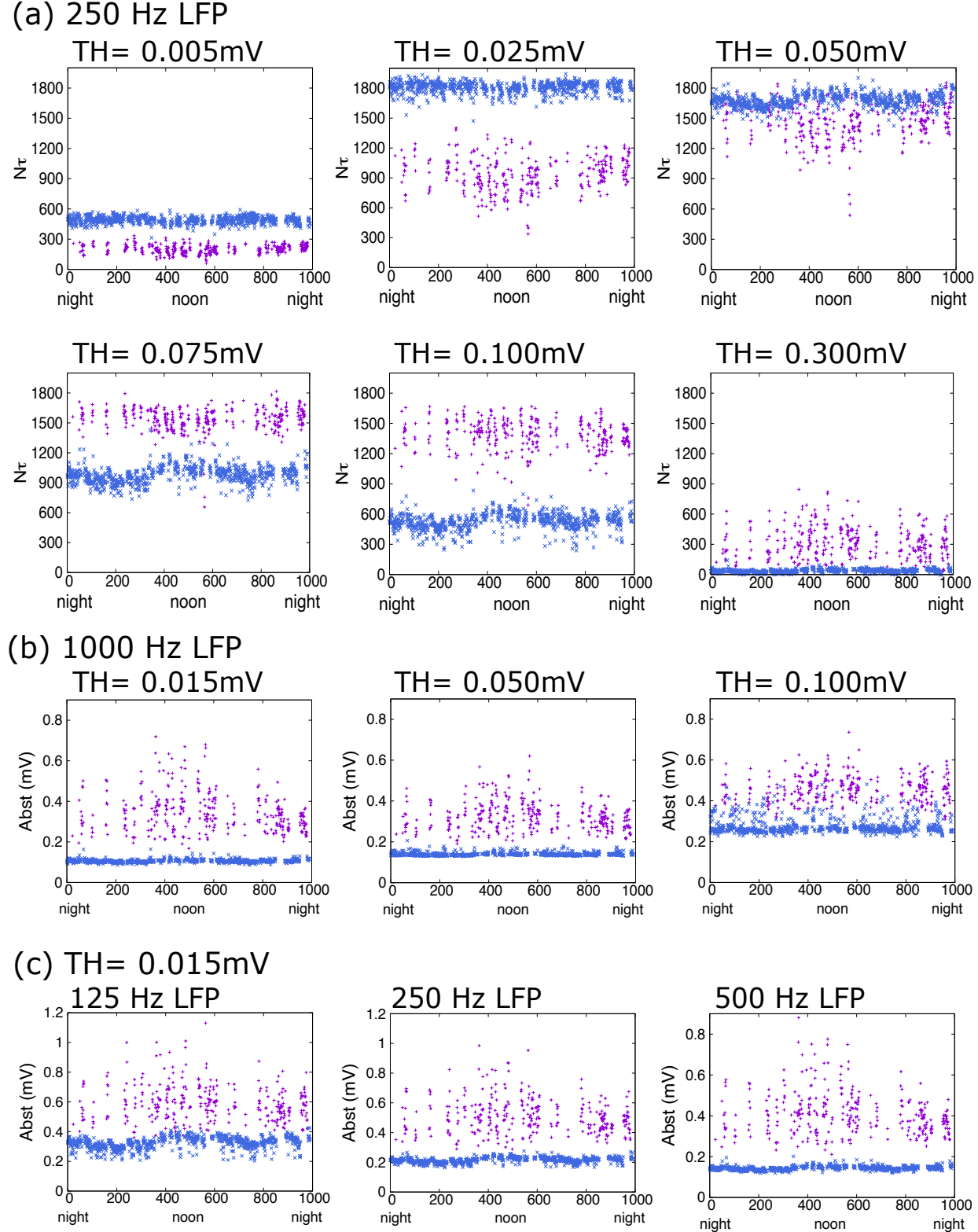

Fig. 3. Example of $N\tau$ and Abst changes by awake (+; n = 307) and sleep (×; n = 684) states with changes in sampling frequency, or with changes in threshold (TH), at M2 in No.1. (a) Distribution of $N\tau$ for trial number on 250Hz LFP at various THs. (b) Distribution of Abst for trial number on 1000Hz LFP at various THs. (c) Distribution of Abst for trial number at TH of 0.015mV on LFP with various sampling rates.

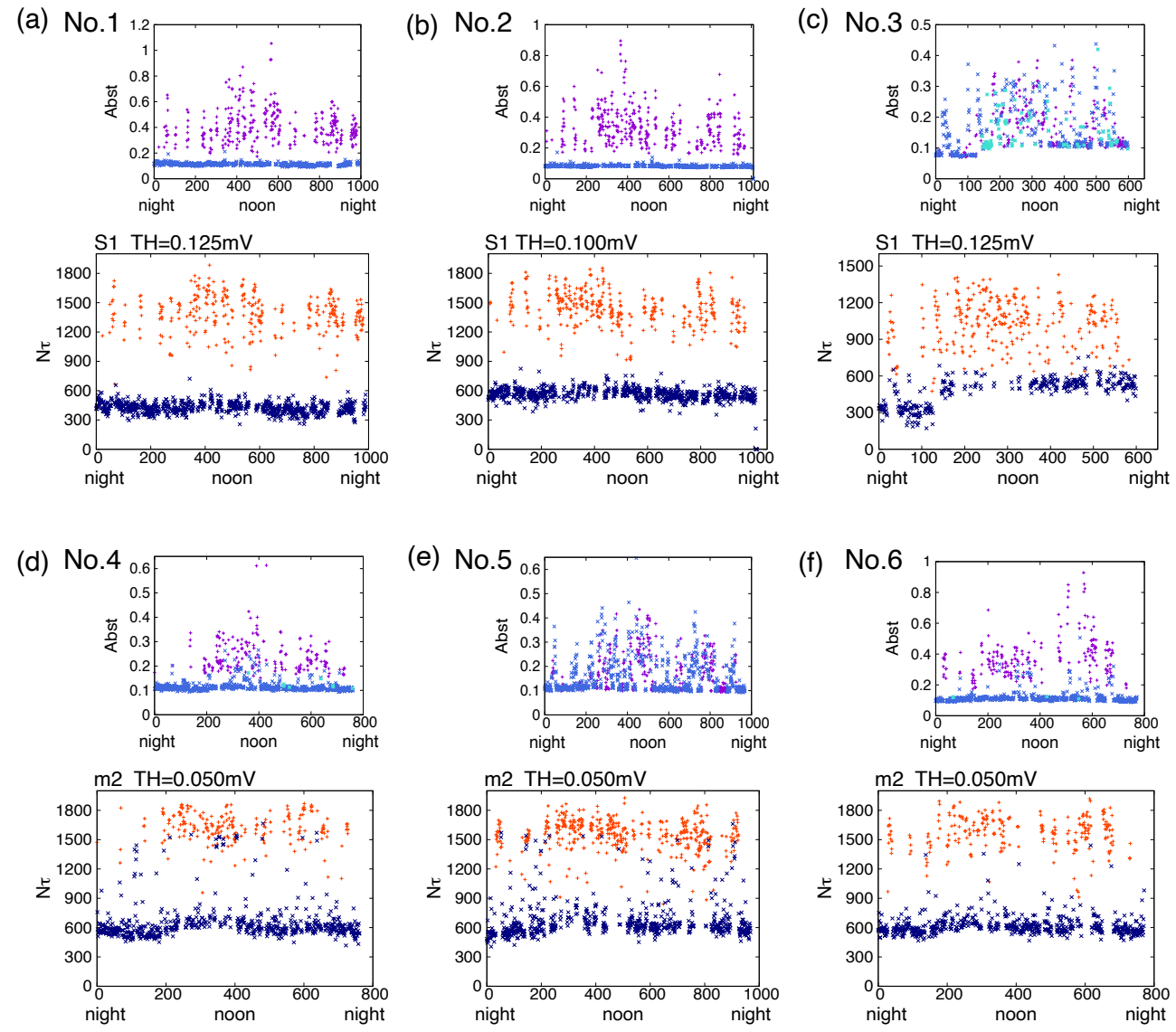

Fig. 4. Hierarchical clustering to Abst on LFP at S1 discriminated two $N\tau$ populations on LFP at M2 and S1. (a-f): Awake state with high Abst and sleep state with low Abst by Ward's method of hierarchical clustering. Scatterplots of Abst from 1000Hz LFP with a threshold (TH) of 0.015mV (a-e: Nos.1-5) and 0.010mV (f: No.6) show classification results of existing method (+: awake, ×: NREM sleep, ✳ : REM sleep) (top). Scatterplots of $N\tau$ from 250Hz LFP at S1 with a TH of 0.125mV (a, c: Nos.1, 3) and 0.100mV (b: No.2) and at M2 with a TH of 0.050mV (d-f: No.4-6) show classification results of Abst (+: awake, ×: sleep) (bottom).

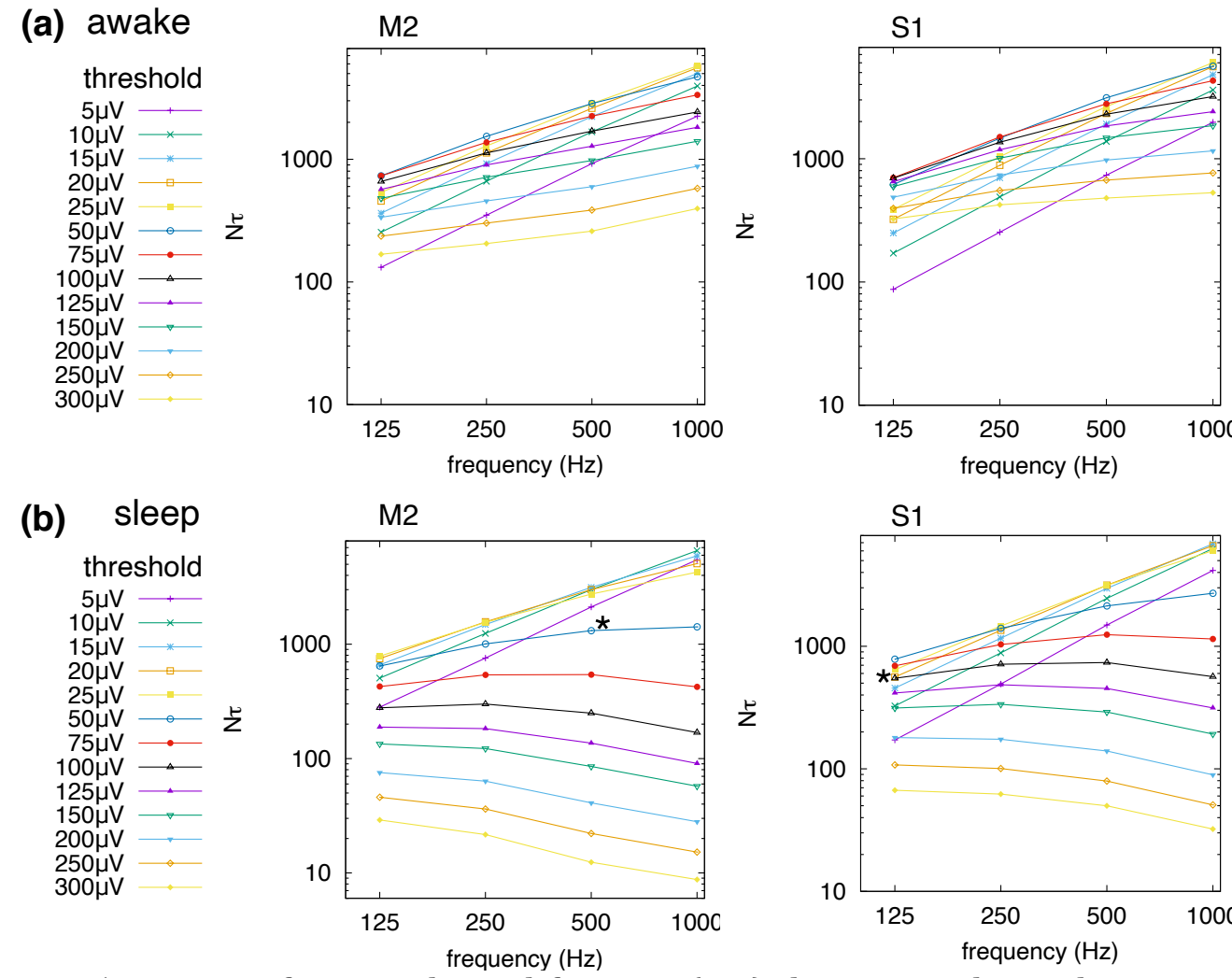

Fig. 5. Positive for apical amplification (AA) during awake and negative for AA during sleep. Relationship between mean $N\tau$ of six mice and sampling rate of LFP with various threshold (TH) at M2 (left) and S1 (right) in awake state (n=1797) (a) and sleep state (n=3309) (b). *: Not significantly different from the value at 1000Hz ($p$>0.05). No mark: significantly different from the values on any other frequencies ($p$< 0.001 in awake at 0.300mV TH, $p$< $10^{-12}$ in others). Statistical significance was assessed by Bonferroni corrected Wilcoxon signed rank test. The state determined by Abst method.

### *C. Discrimination between awake and sleep states using Abst*

Hierarchical clustering using Ward's linkage was performed for 2-state (awake and sleep) discrimination using Abst calculated on 1000Hz LFP at S1 with threshold of 0.015mV (Nos.1-5) or 0.010mV (No.6). The high values of Abst were discriminated into awake state and the low values of Abst into sleep state. These 2 classifications for Abst at S1 were adaptable to discriminating between the two populations of $N\tau$ at both S1 and M2 (Fig.4, Supplementary Fig.1). The number of trials in 2 states by Abst is shown in Fig. 6c.

### *D. Spike-like neural activity occurs only during awake*

At threshold greater than amplitude of microwave, the increase of $N\tau$ with increasing sampling rate suggests the occurrence of spikes (Fig.1). To authenticate whether enhancement of AA occurs consistently only during arousal, we tested whether $N\tau$ decreases in sleep state and increases in awake state with increasing sampling frequency of LFP above a certain threshold. Using the results of clustering to Abst, all trials of six mice were divided into two groups: awake (n=1797) and sleep (n=3309). For each group at M2 and S1, $N\tau$ was compared among 4 sampling frequencies at 13 different thresholds (Fig.5): In awake state, $N\tau$ increased significantly as the sampling frequency increased ($p$<0.001: threshold= 0.300mV at S1, $p$< $10^{-12}$: others). In sleep state, as sampling frequency increased, $N\tau$ decreased significantly at threshold above 0.125mV at M2, and at threshold above 0.200mV at S1 (both $p<10^{-15}$), whereas Nτ increased significantly at threshold below 0.025mV at M2, and at threshold below 0.050mV at S1 ($p<10^{-15}$). These results indicate that AA at M2 and S1 is quiescent throughout sleep state.

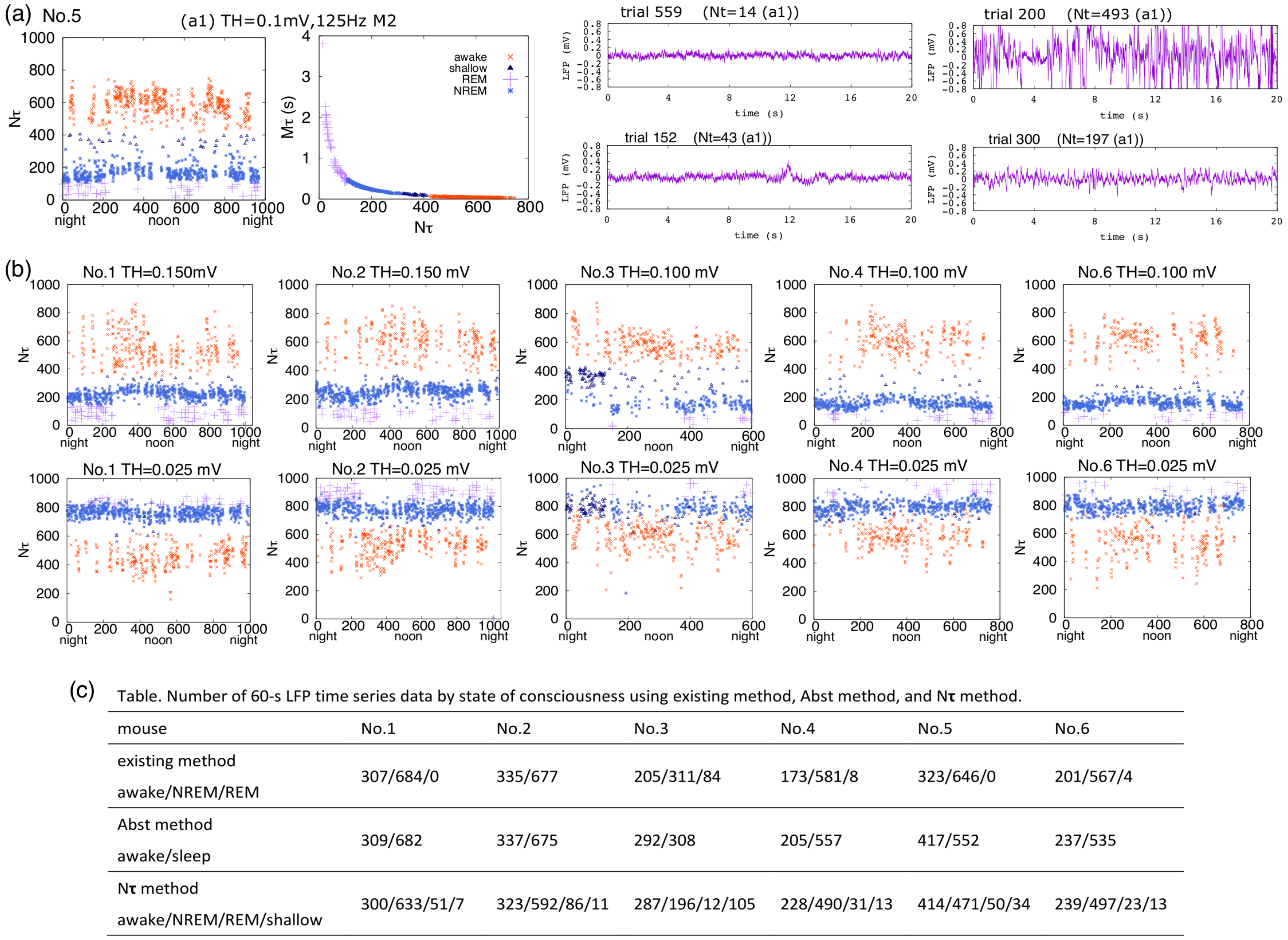

Table. Number of 60-s LFP time series data by state of consciousness using existing method, Abst method, and Nτ method.

| mouse | No.1 | No.2 | No.3 | No.4 | No.5 | No.6 |
|---|---|---|---|---|---|---|
| existing method awake/NREM/REM | 307/684/0 | 335/677 | 205/311/84 | 173/581/8 | 323/646/0 | 201/567/4 |
| Abst method awake/sleep | 309/682 | 337/675 | 292/308 | 205/557 | 417/552 | 237/535 |
| Nτ method awake/NREM/REM/shallow | 300/633/51/7 | 323/592/86/11 | 287/196/12/105 | 228/490/31/13 | 414/471/50/34 | 239/497/23/13 |

Fig. 6. Four-state classification using **Nτ** on 125Hz LFP at M2; awake state (×), shallow sleep (△), REM sleep (+), and NREM sleep (✱). (a) Scatterplots of **Nτ** for trial number (left) and **Mτ** for **Nτ** (second from left) at TH of 0.100mV and 20s LFP waveforms sampling rate at 1000Hz (second from right and right) in No.5. **Nτ**, shown above the diagram, is the value corresponding to the left scatterplots. (b) Scatterplots of **Nτ** for trial number with TH of 0.100-0.150mV (top) and 0.025mV (bottom). (c) Number of trials by state of consciousness using existing method, Abst method, and **Nτ** method.

### *E. Discrimination between awake, NREM sleep, REM sleep, and shallow sleep using Nτ on 125Hz LFP of M2*

Hierarchical clustering was performed in two steps for consciousness discrimination of 4-states using **Nτ** on 125Hz LFP at M2 with threshold of 0.150mV (Nos.1-2) or 0.100mV (Nos.3-6). These thresholds were determined where most **Nτ**s in sleep state are roughly around 200 (Figs.6ab). The first step was that, using Ward's linkage of hierarchical clustering, the entire trials were divided into 2 classes, which were assigned to awake and sleep groups in descending order of **Nτ**. The second step was that, using centroid linkage (No.1, Nos.3-6) or average linkage (No.2) of hierarchical clustering, the trials of the sleep group were further divided into 3 classes, which were assigned to shallow sleep, NREM sleep, and REM sleep groups in descending order of **Nτ**. As shown in the waveform of Fig.6a, the smaller **Nτ** is, the more typical REM waveform is; as **Nτ** increases, it is accompanied by slow waves and then spikes. At threshold of 0.100mV or 0.150mV, **Nτ** was significantly lower and **Mτ** was significantly higher in REM, NREM, shallow sleep states, and awake state in that order ($p<10^{-15}$) (Fig.6ab). Those classifications were consistence for Nt subpopulations at threshold of 0.025mV (Fig.6b).

### *F. Frequency analysis for state of consciousness*

To evaluate LFP by spectral edge frequency 95 % (SEF95), we performed power spectral analysis using Fast Fourier transform (Supplementary table 2). In all subjects, SEF95 did not discriminate between the 4 states. In Nos.1 and 2, where 2 classifications (awake and sleep) were in almost perfect agreement between new method and existing method, SEF95s at M2 in awake state were very high with small variations compared to the others: 424 (411, 436) (median (1st Qu, 3rd Qu)), 420 (403, 433), 95 (23, 306), 72 (21, 291), 34 (17, 130), and 33 (18, 182) (Hz) in 1000Hz LFP of Nos. 1 (n=300), 2 (n=323), 3 (n=287), 4 (n=228), 5 (n=414), and 6 (n=239), respectively. In the other four cases, there was no significant difference in SEF95 between awake and REM sleep states. In two of the four cases, there was no significant difference in SEF95 between awake and NREM sleep states. Thus, **τ** and **burst** method showed robust state discrimination performance even for LFP data with large frequency variation.

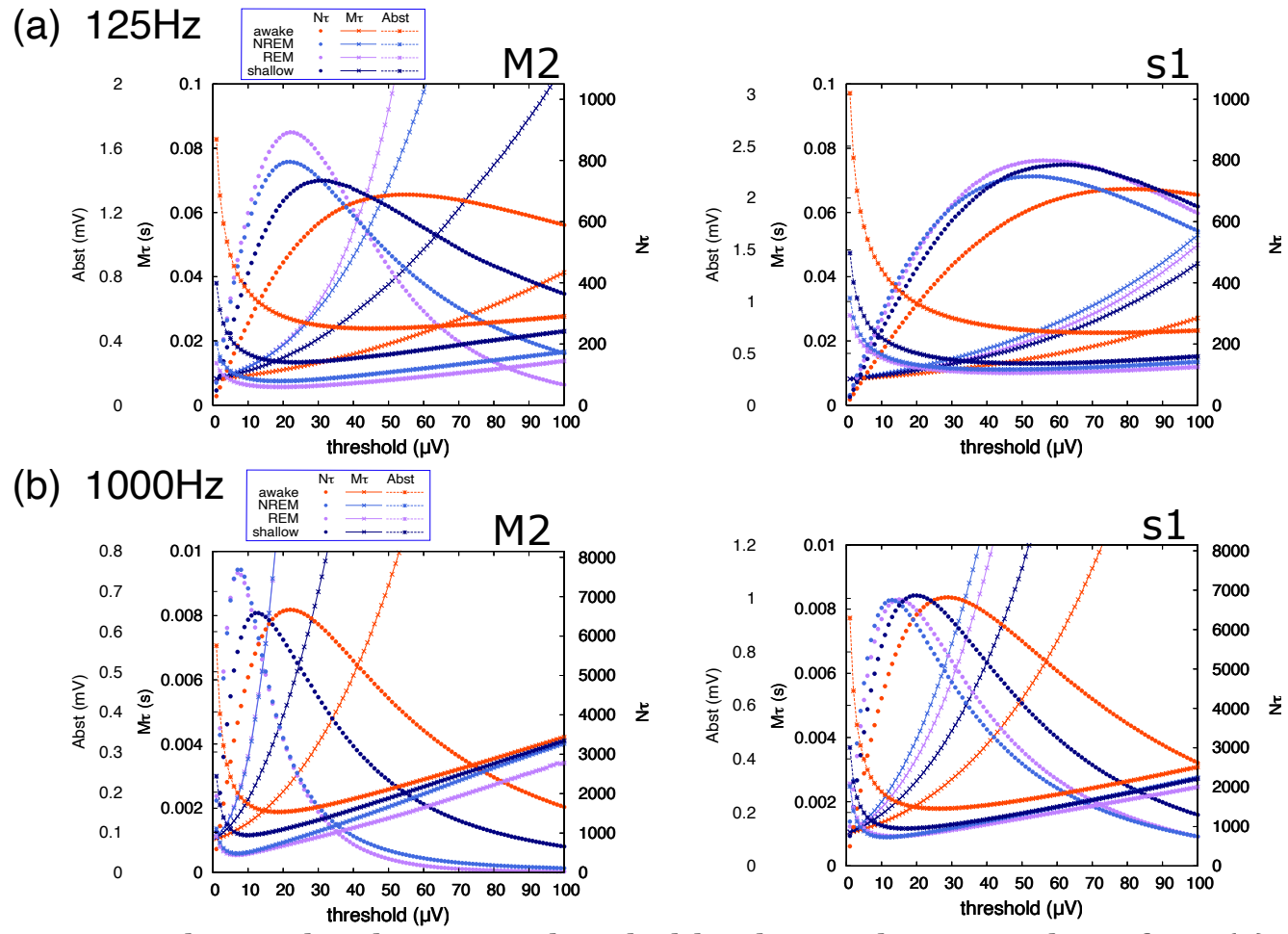

Fig. 7. Relationship between threshold value and mean value of N**τ** (•), M**τ** (×), or Abst (*) on LFP sampling rate of 125Hz (a) and 1000Hz (b) at M2 (left) and S1 (right) in awake state (colored orange), NREM sleep state (colored blue), REM sleep state (colored purple), and shallow sleep state (colored navy blue) in No.5.

## G. *Threshold to maximize Nτ is proportional to amplitude of microwaves*

For purpose of exploring the conditions that allow for comparison and the principle of **τ**, we studied the threshold at which N**τ** is the maximum value (THnt) in each state, together with the indices at THnt, using mean indices at thresholds for each 0.001mV interval (Figs.7, Supplementary Figs.2-5 and table3).

At THnt, mean M**τ** was 2.5 to 3.0 times the sampling rate of LFP in any state (Fig.8a). By observing enlarged waveforms, we confirmed that there were fluctuations in each sample at any rate of LFP (ex Fig.1def). An example of THnt event where N**τ** goes from increasing to decreasing with increasing threshold is shown in Fig.8b. In order for N**τ** to decrease, separate **τ** must be coupled (Fig.8b). These results suggest that THnt is proportional to the amplitude of the microwave. In addition, since the conditions at THnt are aligned with M**τ**, it is possible to compared the indices at THnt between the states.

THnt and mean Abst at THnt were significantly different among the 4 states, and between M2 and S1 at any state (Fig. 8a). Abst at THnt was higher in awake state, shallow sleep, NREM sleep, and REM sleep states, in that order ($p<10^{-7}$). The difference in mean Abst at THnt between NREM and REM sleep states decreased with increasing frequency, and then both mean Abst at THnt values became almost equal on 1000Hz LFP. THnt was higher in awake state, shallow sleep, REM sleep, and NREM sleep states, in that order ($p<10^{-5}$). The difference in THnt between REM and NREM sleeps was greater at S1 than at M2. Mean THnt of 4 different frequencies in REM and NREM sleeps was 34.0mcV and 29.8mcV at S1, and 19.0mcV and 18.3mcV at M2.

As for the discrimination between NREM and REM sleep states, N**τ** method was derived from the characteristics of waveform on 125Hz data, such as slow wave for NREM and p values from exact Wilcoxon signed rank test with Holm correction (n=48).

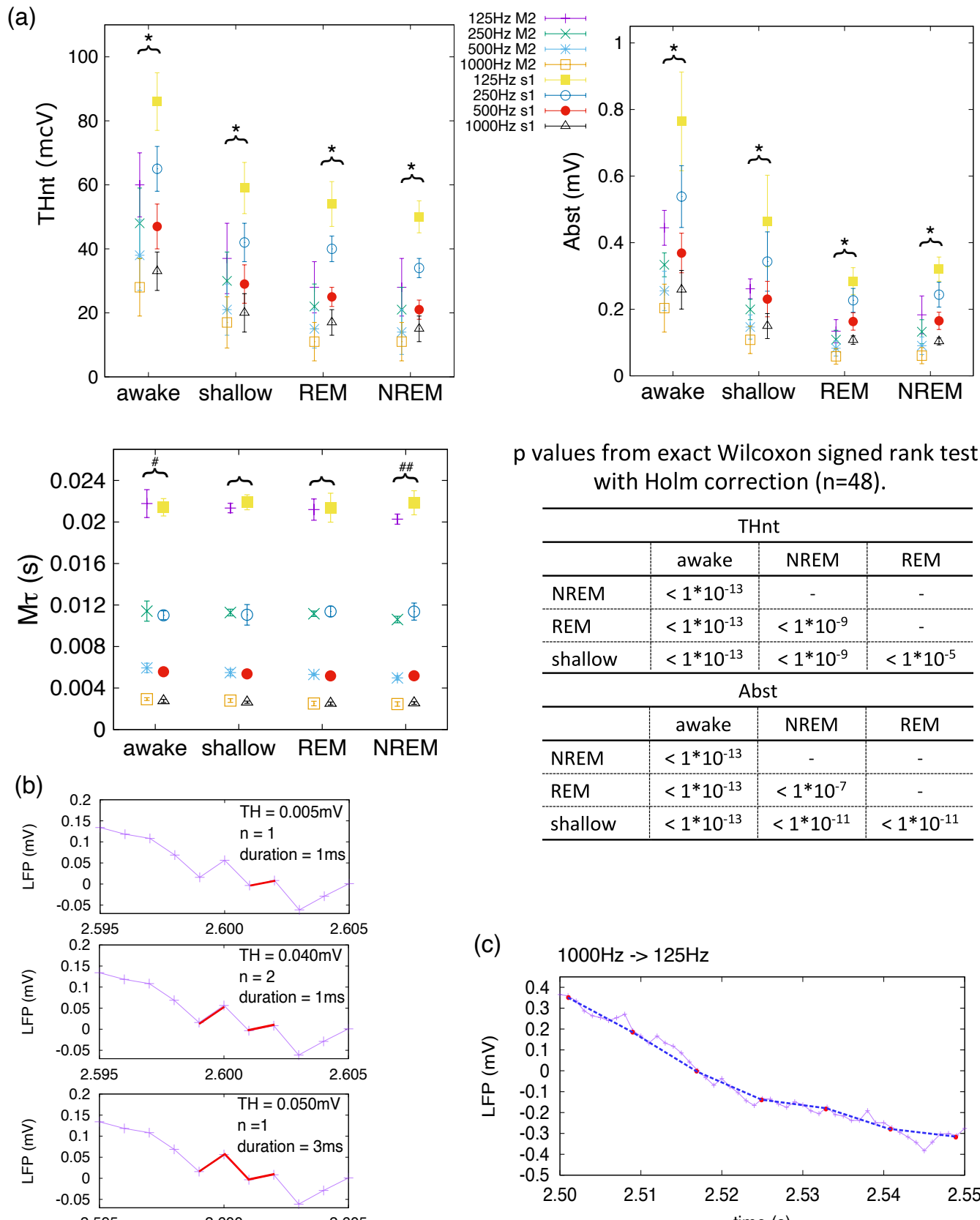

| THnt | | | |
|---|---|---|---|
| | awake | NREM | REM |
| NREM | $< 1*10^{-13}$ | - | - |
| REM | $< 1*10^{-13}$ | $< 1*10^{-9}$ | - |
| shallow | $< 1*10^{-13}$ | $< 1*10^{-9}$ | $< 1*10^{-5}$ |

| Abst | | | |
|---|---|---|---|
| | awake | NREM | REM |
| NREM | $< 1*10^{-13}$ | - | - |
| REM | $< 1*10^{-13}$ | $< 1*10^{-7}$ | - |
| shallow | $< 1*10^{-13}$ | $< 1*10^{-11}$ | $< 1*10^{-11}$ |

Fig. 8. (a)Mean and SD of indices at threshold for maximizing mean N**τ** (THnt) in awake, shallow sleep, REM sleep, and NREM sleep states. Sampling rate of 125Hz, 250Hz, 500Hz, 1000Hz in LFP are coded by −, ×, ✱, □ at M2 (n = 24), and by ■,○,●,Δ at S1 (n = 24), respectively. Top: THnt for state of consciousness (left), Abst for state of consciousness (right). Middle: M**τ** for state of consciousness (left), p values from exact Wilcoxon signed rank test with Holm correction to compare indices between states of consciousness (right). Statistical significance among M2 and S1 was assessed by exact Wilcoxon signed rank test; *$p<10^{-5}$, ##p<0.001, #p<0.05. (b) Example of fluctuations in N**τ** with M**τ** as the threshold (TH) rises on LFP sampling rate at 1000Hz. (c) Two types LFP waveforms sampled at 1000Hz (purple line-points) and at 125Hz (blue line and red points). When sampling rate is reduced, microwaves on the slope disappear.

low-amplitude for REM. As shown in Fig.8c, on slow wave, the microwaves expressed on 1000Hz LFP disappear on 125Hz LFP. Hence, on 125Hz LFP, N**τ** at THnt (near 0.025mV) was larger in REM sleep state than in NREM sleep state (Fig.6b), whereas there was no different at 1000Hz. In addition, the lower N**τ** at threshold above 0.100mV was derived from the lower amplitude wave in REM than in NREM, as also supported by M**τ** (Fig.6). As shown in Fig.1ef, with increasing frequency, the waves in NREM and REM sleep states fluctuates and their amplitudes are subdivided. Indeed, the difference of amplitude between the wave of the two states is most obvious at 125Hz LFP (Fig.8a, Supplementary Figs.2-5).

## IV. Discussion

In this study, we have demonstrated the effectiveness of our original indices in discrimination of consciousness state in mice and shown that amplitude of microwaves differed depending on the state of consciousness. The pattern of changes in indices in respect to threshold variation was almost the same in all cases, despite the variety of SEF95 for each subject. Our hypothesis was inferred from consciousness level-specific waveform morphology based on reports that arousal enhances AA. Consistent with that hypothesized, Abst showed clearly greater values and variability in awake state than in sleep state. We also demonstrated persistence of AA rest during sleep by showing that N**τ** at threshold above 0.125mV or 0.200mV decreased in sleep state and increased in awake state with increasing sampling frequency. On the other hand, because M**τ** at THnt was aligned at 2.5 to 3.0 times the sampling interval, THnt was proportional to amplitude of the microwave. Indeed, the enlarged waveform diagram confirms that each sample can fluctuate regardless of the sampling rate. Therefore, THnt would indicate instability of the electrode ground field, not electrical activity of neuron. Because THnt was significantly higher in order of awake, shallow sleep, REM sleep, and NREM sleep states, THnt was correlated with the level of consciousness. Furthermore, if the smaller difference in THnt between REM and NREM in M2 was due to muscle atonia during REM sleep, THnt would reflect the oscillations produced by neural activity.

As far as we know, the effects of environment surrounding the electrodes, such as astrocytes, fluid flow, and magnetic fields, on baseline of LFP waveforms have not yet been properly investigated. Recent reports have shown that adrenergic signaling are associated with elevated astrocyte $Ca^{2+}$ levels and cerebrospinal fluid dynamics [23] [24]. We speculate that the microwaves of LFPs reflect these environments around the electrodes, and further studies are needed to elucidate the physical mechanisms of THnt, amplitude of baseline.

As shown on the scatterplot in No.3 (Figs.4c,6b), remarkable collective deviation of indices in the sleep state at both M2 and S1 was observed in trials with the number less than 150. The cause has not been investigated sufficiently and has not been identified. In all subjects, the appearance of REM sleep-like waveforms was scarce during the bright hours. (Fig.6ab). This may be related to the suppression of the appearance of REM sleep-like EEG by blue light [25].

## V. Conclusion

The original **τ** and **burst** indices developed in dogs was useful for discriminating the conscious state in mice using LFP at layer 5 of S1 and M2. Furthermore, the microwave that we focused on as **τ** were shown to originate from the fluctuations of the electrode ground field, and their amplitude (THnt) was correlated with the state of consciousness. Further studies are needed to identify the factors involved in amplitude of microwaves.

### Data Access Statement

All data and program code are available upon reasonable request to the corresponding author.

### Acknowledgment

The authors would like to acknowledge Masanori Murayama and and his colleagues at Center for Brain Science RIKEN for collecting data and conception of this work. We also gratefully acknowledge Satoshi Hagihira at Kansai Medical University, and Jun Tamura, Takaharu Itami, Tomohito Ishizuka and the students at Rakuno Gakuen University, who supported anesthesia study in dogs that resulted in the development of **τ** and **burst** indices.

# Supplementary figures

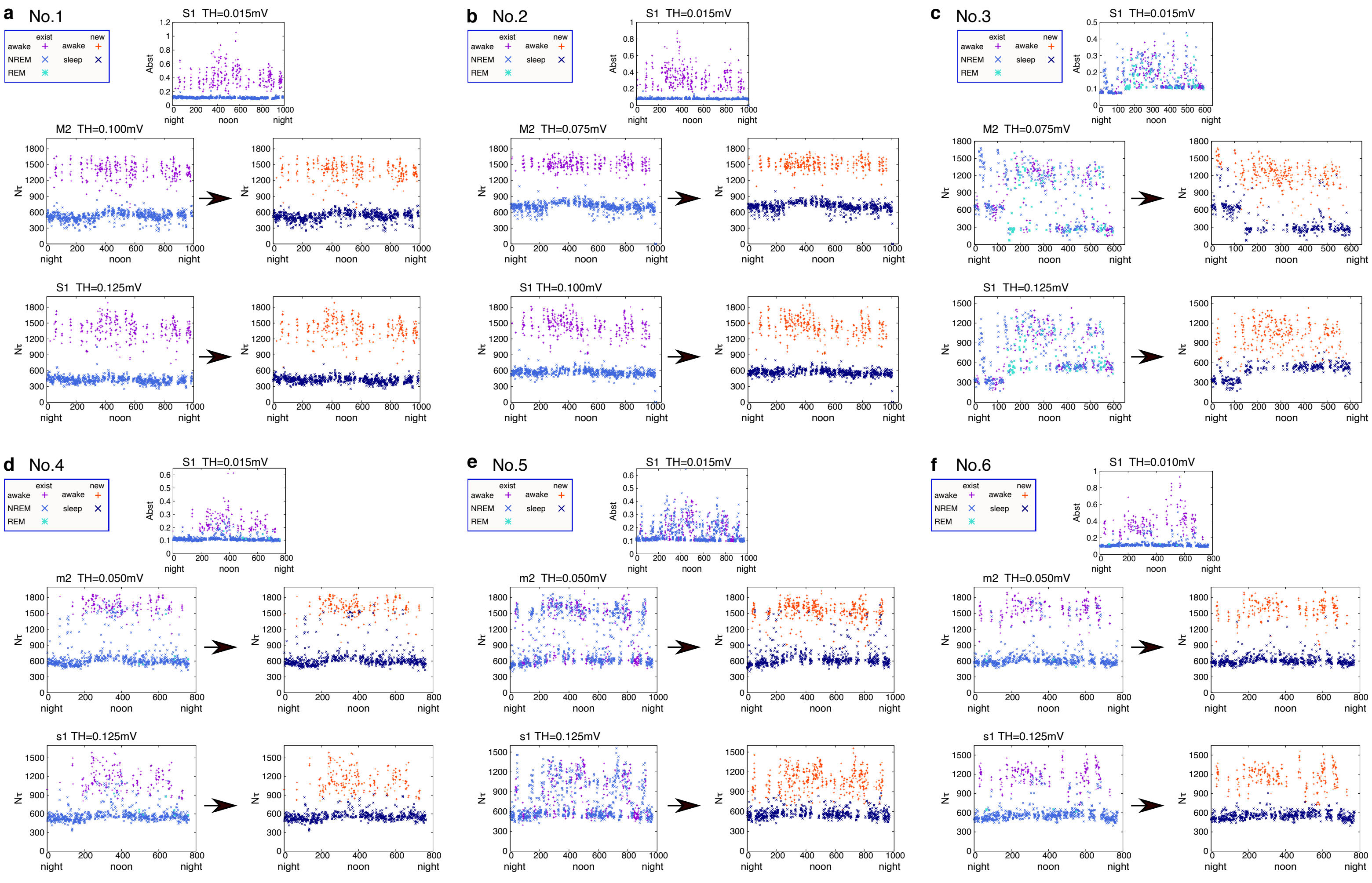

**Supplementary Figure 1 Hierarchical clustering to Abst discriminated 2 populations of Nτ.** Awake state with high Abst and sleep state with low Abst by Ward's method of hierarchical clustering. Abst from 1000 Hz LFP with threshold (TH) of 0.015mV (a-e: Nos.1-5) and 0.010mV (f: No.6) were used for clustering (top). Scatterplots of N**τ** from 250Hz LFP show classification results using existing method (+: awake, ×: NREM sleep, * ; REM sleep) in left and Abst method (+: awake, ×: sleep) in right. Those in M2 at TH of 0.100mV (a: No.1), 0.075mV (b, c: Nos.2, 3) or 0.050mV (d-f: Nos. 4-6) are shown in middle row and those in S1 at TH of 0.125mV (Nos. 1, 3-5) or 0.100mV (No.2) is shown in bottom row.

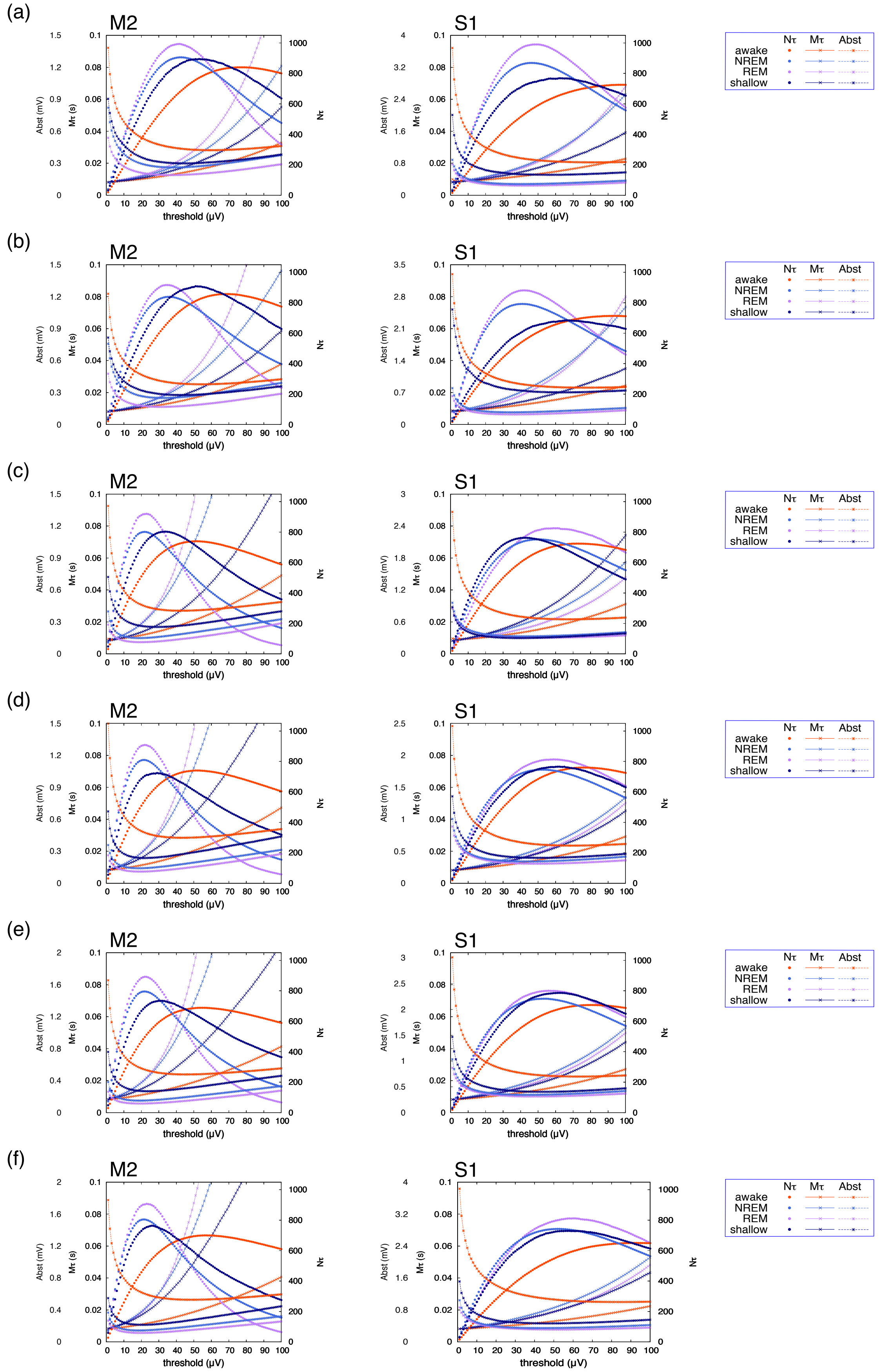

**Supplementary Figure 2**

Relationship between threshold value and N**τ** (•), M**τ** (×), or Abst (*) on local field potential sampling rate of 125Hz at secondary motor cortex (left) and primary somatosensory cortex (right) in awake state (colored orange), NREM sleep state (colored blue), REM sleep state (colored purple), and shallow sleep state (colored navy blue) in Nos.1 (a), 2 (b), 3 (c), 4 (d), 5 (e), and 6 (f).

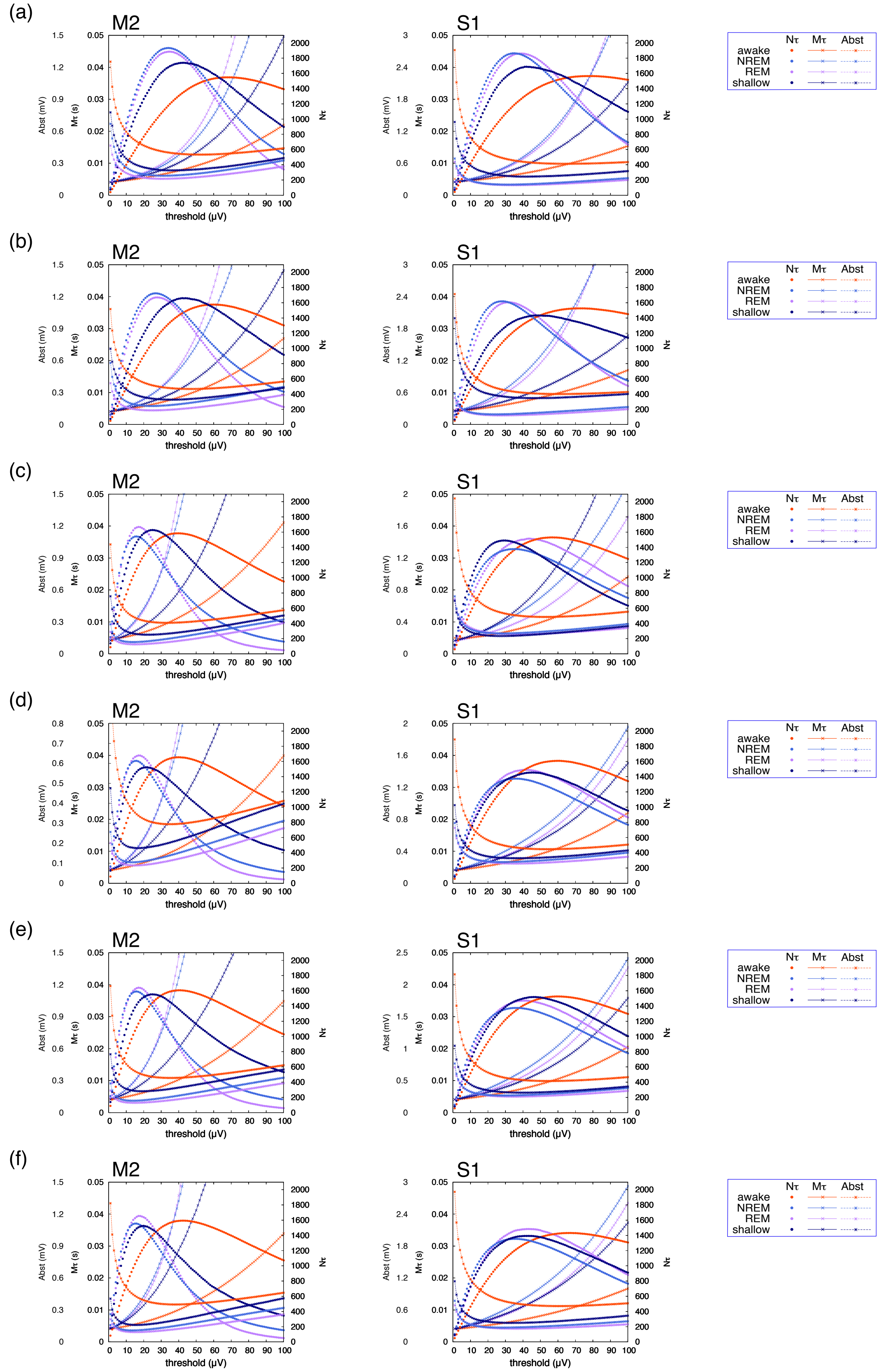

**Supplementary Figure 3**

Relationship between threshold value and N**τ** (•), M**τ** (×), or Abst (*) on local field potential sampling rate of 250 Hz at secondary motor cortex (left) and primary somatosensory cortex (right) in awake state (colored orange), NREM sleep state (colored blue), REM sleep state (colored purple), and shallow sleep state (colored navy blue) in Nos.1 (a), 2 (b), 3 (c), 4 (d), 5 (e), and 6 (f).

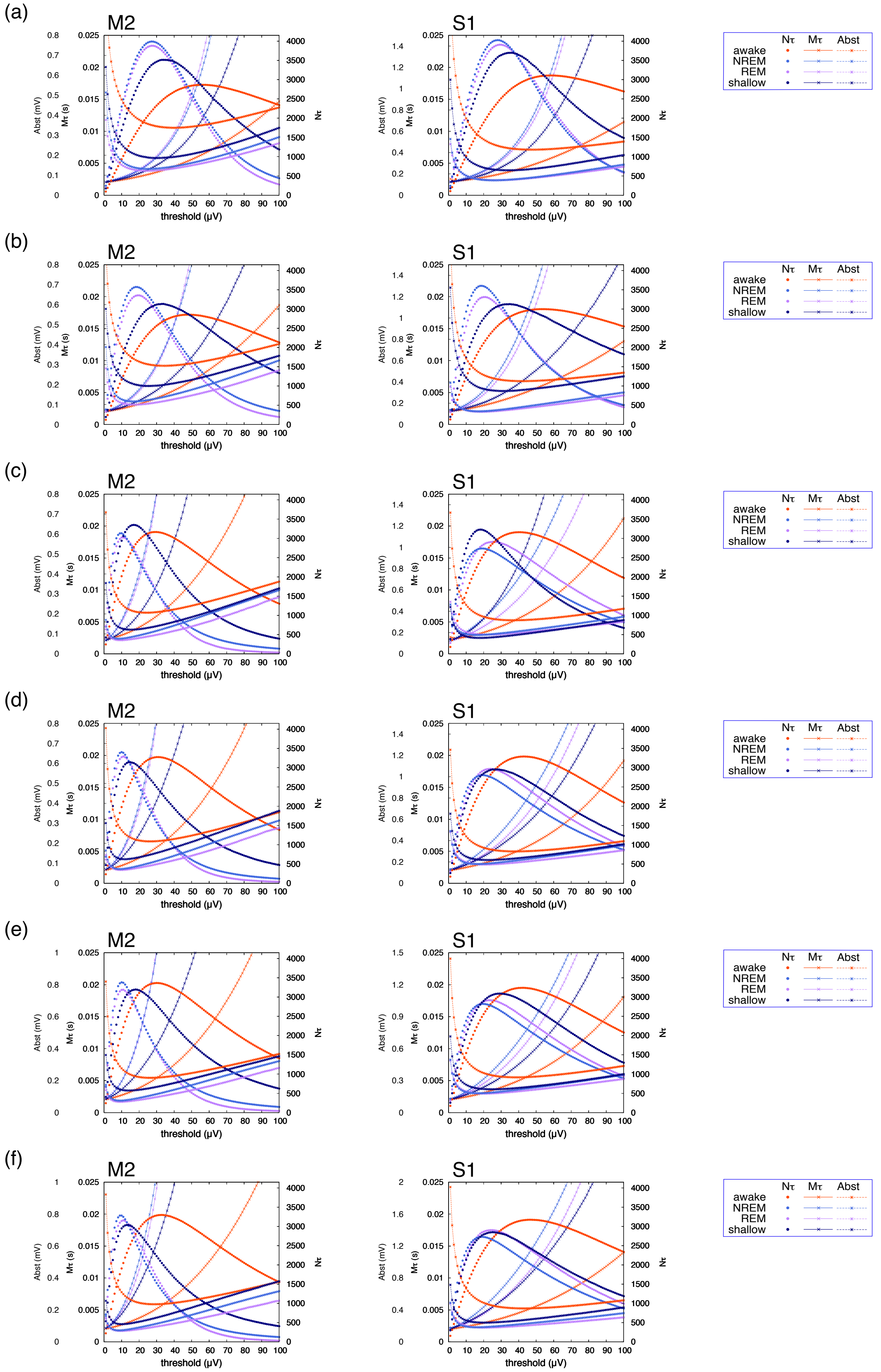

**Supplementary Figure 4**

Relationship between threshold value and N$\boldsymbol{\tau}$ (•), M$\boldsymbol{\tau}$ (×), or Abst (*) on local field potential sampling rate of 500 Hz at secondary motor cortex (left) and primary somatosensory cortex (right) in awake state (colored orange), NREM sleep state (colored blue), REM sleep state (colored purple), and shallow sleep state (colored navy blue) in Nos.1 (a), 2 (b), 3 (c), 4 (d), 5 (e), and 6 (f).

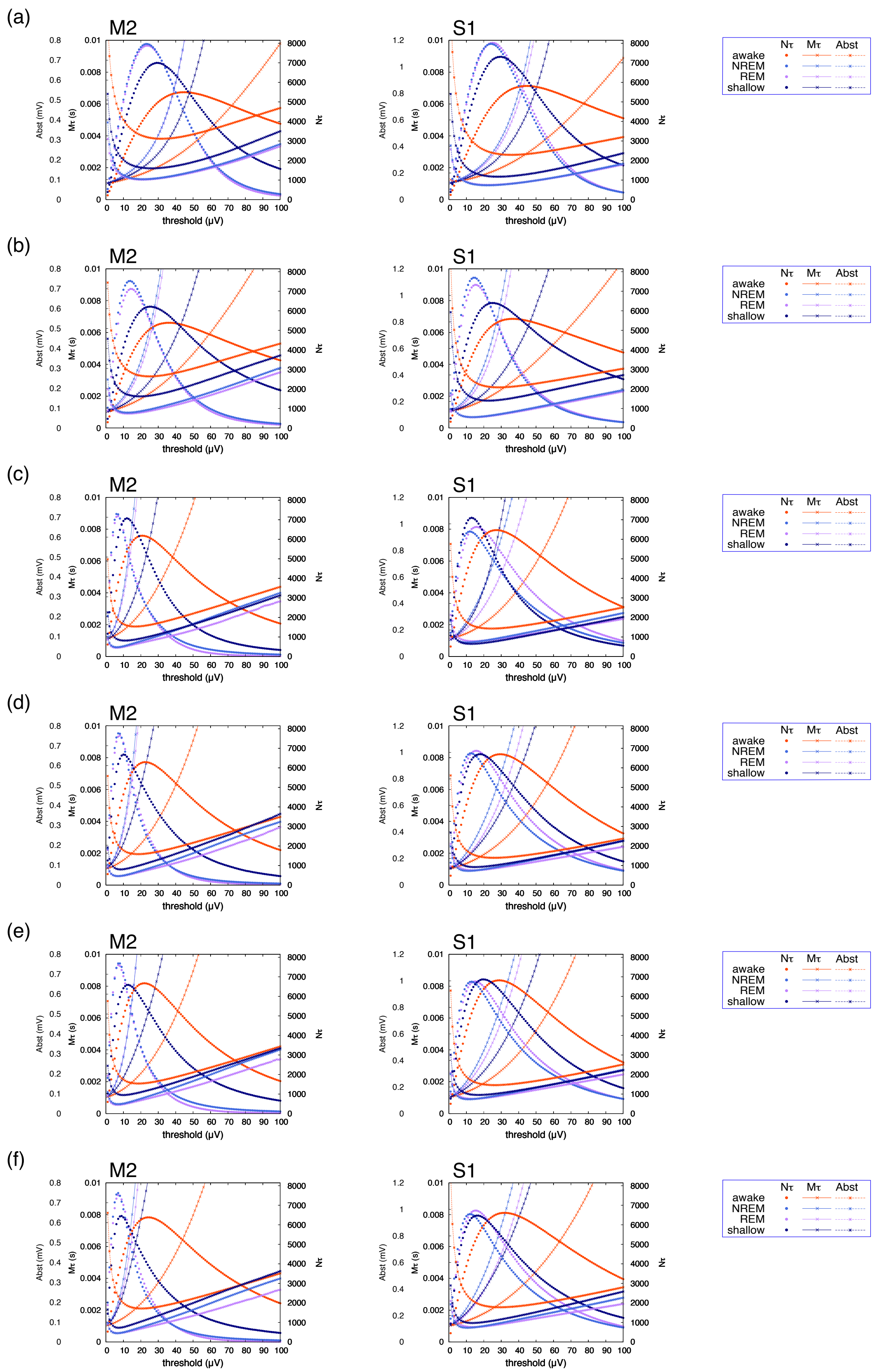

**Supplementary Figure 5**

Relationship between threshold value and N**τ** (•), M**τ** (×), or Abst (*) on local field potential sampling rate of 1000 Hz at secondary motor cortex (left) and primary somatosensory cortex (right) in awake state (colored orange), NREM sleep state (colored blue), REM sleep state (colored purple), and shallow sleep state (colored navy blue) in Nos.1 (a), 2 (b), 3 (c), 4 (d), 5 (e), and 6 (f).

**Supplementary Table**

| | state | sampling frequency | | | |
|---|---|---|---|---|---|
| | | 125Hz | 250Hz | 500Hz | 1000Hz |
| Nτ (/60s) | awake | 800±53 | 1391±144* | 2325±339*" | 3832±727*"# |
| | NREM | 464±55 | 524±85* | 425±109*" | 266±163*"# |
| Mτ (ms) | awake | 33.3±7.1 | 22.5±6.2* | 15.1±5.0*" | 10.2±4.0*"# |
| | NREM | 85.7±21.3 | 95.6±24.3* | 138.3±38.7*" | 276.3±121.0*"# |
| Abst (mV) | awake | 0.457±0.063 | 0.434±0.060* | 0.435±0.060* | 0.456±0.063"# |
| | NREM | 0.370±0.034 | 0.322±0.022* | 0.289±0.020*" | 0.277±0.040*"# |

Supplementary Table 1 Mean ± SD of number of τ (Nτ), mean duration of τ (Mτ), and mean amplitude of burst (Abst) at a threshold of 0.100mV for every 60s of data during 20 hours of awake (n=302) and NREM (n=645) in No.1. State of consciousness was determined by the existing method. $p<1*10^{-15}$; *, ", #: Significantly difference from the values at 125Hz, 250Hz, and 500Hz, respectively (Bonferroni corrected Wilcoxon signed rank test).

Supplementary table 2 | Median (1st Qu, 3rd Qu) of spectral edge frequency 95% (SEF95) (Hz) on LFP sampled at 125Hz and 1000 Hz at secondary motor cortex

| | id | SEF95 (Hz) at M2 | | | | SEF95 (Hz) at S1 | | | |
|---|---|---|---|---|---|---|---|---|---|
| | | awake | NREM | REM | shallow | awake | NREM | REM | shallow |
| 125 Hz | No.1 | 56 (55, 57) | 19 (17, 22)* | 40 (38, 42)*# | 54 (18, 54) | 21 (15, 32) | 24(21, 26) | 35 (29, 40)*# | 21 (15, 28)' |
| | No.2 | 55 (53, 56) | 16 (15, 19)* | 37 (27, 40)*# | 53 (50, 54)#" | 20 (14, 36) | 22 (21, 24) | 32 (25, 37)+# | 18 (15, 31) |
| | No.3 | 26 (15, 41) | 21 (17, 23)* | 24 (22, 35) | 17 (15, 21)+¦ | 21 (17, 36) | 26 (24, 28)* | 34(30, 38)+# | 24 (22, 26)#" |
| | No.4 | 26 (15, 39) | 22 (21, 25) | 37 (31, 42)+# | 21 (17, 31)' | 28 (19, 39) | 26 (24, 28) | 32 (28, 37)# | 28 (22, 32) |
| | No.5 | 18 (13,34) | 22 (21, 24)* | 35 (26, 43)*# | 22 (17, 34)" | 22 (16, 35) | 26 (24, 28)* | 30 (25, 37)*# | 26 (19, 33) |
| | No.6 | 19 (14,32) | 21 (20, 23)+ | 36 (28, 42)*# | 21 (21, 27)' | 21 (16, 25) | 25 (23, 27)* | 35 (29, 37)*# | 21 (17, 23)¦' |
| 1000 Hz | No.1 | 424 (411, 436) | 23 (20, 34)* | 166 (127, 218)*# | 399 (19, 412)+ | 138 (23, 270) | 40 (32, 53)* | 102 (76, 138)# | 27 (22, 32)' |
| | No.2 | 420 (403, 433) | 18 (16, 21)* | 63 (36, 88)*# | 416 (375, 423)#" | 79 (19, 283) | 27 (24, 30)* | 51 (31, 60)# | 24 (18, 213) |
| | No.3 | 95 (23, 306) | 23 (20, 28)* | 34 (33, 63)# | 19 (17, 28)*¦' | 29 (19, 81) | 29 (27, 31) | 39 (33, 53)¦ | 26 (24, 32)¦' |
| | No.4 | 72 (21, 291) | 26 (23, 31)* | 66 (53, 86)# | 34 (23, 46)' | 44 (23, 90) | 29 (27, 32)* | 37(31, 51)# | 42 (28, 54) |
| | No.5 | 34 (17, 130) | 25 (23, 30) | 56 (30, 72)# | 34 (23, 54) | 27 (19, 48) | 29 (26, 32) | 35 (28, 47)# | 32 (23, 48) |
| | No.6 | 33 (18, 182) | 24 (22, 27)+ | 61 (38, 70)# | 28 (22, 62) | 23 (18, 35) | 27 (25, 30)* | 42 (31, 45)*# | 23 (20, 29)' |

(M2) and primary somatosensory cortex (S1) in awake state, and 3 states of sleep (NREM, REM, and shallow) in each mouse. The state of consciousness was determined by N**τ** method.

p <0.001; *, #, " :significant difference from the value of awake, NREM, REM, respectively,

p <0.05; +, ¦, ' :significant difference from the value of awake, NREM, REM, respectively (Bonferroni corrected Wilcoxon rank sum test)

| | (Hz) | awake (n=1791) | | shallow (n=183) | | REM (n=258) | | NREM (n=2879) | |
|---|---|---|---|---|---|---|---|---|---|
| | | M2 | S1 | M2 | S1 | M2 | S1 | M2 | S1 |
| $N\tau$ | 125 | 761±103 | 713±107 | 792±95 | 761±77 | 926±130 | 870±142 | 833±56 | 783±65 |
| | 250 | 1592±137 | 1531±159 | 1605±133 | 1490±126 | 1707±239 | 1601±259 | 1687±158 | 1529±211 |
| | 500 | 3137±329 | 3159±253 | 3277±282 | 3168±335 | 3386±533 | 3256±577 | 3524±332 | 3223±551 |
| | 1000 | 6077±766 | 6330±689 | 6856±709 | 6950±620 | 7470±1007 | 7214±1082 | 7690±444 | 7152±667 |
| $M\tau$ | 125 | 21.8±4.2 | 21.4±4.9 | 21.2±2.6 | 21.8±3.4 | 20.7±2.9 | 20.6±3.3 | 20.2±1.3 | 21.7±1.8 |
| | 250 | 11.4±2.4 | 11.0±2.5 | 11.0±1.3 | 11.0±1.8 | 11.1±1.6 | 11.2±1.8 | 10.5±0.8 | 11.2±1.1 |
| | 500 | 5.9±1.2 | 5.6±1.2 | 5.4±0.8 | 5.2±0.8 | 5.3±0.8 | 5.1±0.8 | 5.0±0.4 | 5.2±0.4 |
| | 1000 | 2.9±0.6 | 2.8±0.6 | 2.7±0.5 | 2.6±0.4 | 2.5±0.4 | 2.5±0.4 | 2.5±0.3 | 2.6±0.3 |
| Abst | 125 | 0.444±0.136 | 0.761±0.377 | 0.258±0.120 | 0.371±0.176 | 0.146±0.040 | 0.268±0.055 | 0.193±0.058 | 0.315±0.050 |
| | 250 | 0.334±0.088 | 0.537±0.232 | 0.189±0.074 | 0.278±0.124 | 0.118±0.032 | 0.212±0.045 | 0.139±0.037 | 0.238±0.044 |
| | 500 | 0.256±0.077 | 0.369±0.129 | 0.132±0.050 | 0.188±0.079 | 0.090±0.026 | 0.152±0.032 | 0.095±0.027 | 0.161±0.030 |
| | 1000 | 0.204±0.083 | 0.259±0.089 | 0.091±0.038 | 0.120±0.048 | 0.065±0.024 | 0.101±0.019 | 0.065±0.025 | 0.103±0.014 |

Supplementary Table 4 | Mean ± SD of new indexes with the threshold for maximizing $N\tau$ in each state of consciousness at secondary motor cortex (M2) and primary somatosensory cortex (S1) in 6 mice.